# Tunable terahertz emission from $Bi_2Sr_2CaCu_2O_{8+\delta}$ mesa devices

T. M. Benseman, A. E. Koshelev, K. E. Gray, W.-K. Kwok, U. Welp
Materials Science Division, Argonne National Laboratory, Argonne IL 60439

K. Kadowaki, M. Tachiki
Institute for Materials Science, University of Tsukuba, Ibaraki 305-8753, Japan

T. Yamamoto
Semiconductor Analysis and Radiation Effects Group, Japan Atomic Energy Agency,
1233 Watanuki-machi Takasaki Gunma 370-1292 Japan

We have measured coherent terahertz emission spectra from $Bi_2Sr_2CaCu_2O_{8+\delta}$ mesa devices, as function of temperature and mesa bias voltage. The emission frequency is found to be tunable by up to 12 % by varying the temperature and bias voltage. We attribute the appearance of tunability to asymmetric boundaries at the top and bottom, and the non-rectangular cross-section of the mesas. This interpretation is consistent with numerical simulations of the dynamics of intrinsic Josephson junctions in the mesa. Easily tunable emission frequency may have important implications for the design of terahertz devices based on stacked intrinsic Josephson junctions.

Josephson junctions naturally convert a dc-voltage into high-frequency electromagnetic oscillations at Josephson frequency $f_J = V_J/\Phi_0$, that is, 1 mV corresponds to 0.482 THz [1]. Here, $\Phi_0$ is the flux quantum and $V_J$ is the dc-voltage across the junction. Recently, we have demonstrated [2] that the intrinsic Josephson junctions (IJJ) [3] in the highly anisotropic high-temperature superconductor $Bi_2Sr_2CaCu_2O_{8+\delta}$ (Bi-2212) can be induced to emit coherent continuous-wave off-chip radiation in the THz frequency range. The samples were designed in such a way that an electromagnetic cavity resonance synchronizes a large number of intrinsic junctions into a macroscopic coherent state [4] enabling the emission of THz radiation with powers up to 5 µW [5, 6] and frequencies up to 0.85 THz [2]. Far-field radiation from Bi-2212 cavities with rectangular [2], square and disk [7, 8] shape has been observed. It is possible that coupling to the resonance mode is facilitated by the formation of a dynamic phase kink state that was found in numerical simulations [9, 10] or by the appearance of bound fluxon – antifluxon pairs [11].

For a long rectangular cavity with width $w$ the radiating cavity mode has frequency $f = c/2w$, where $c$ is the speed of the electromagnetic waves in BSCCO. In the uniform fully synchronized state the electromagnetic fields in the mesa do not depend on the c-axis coordinate and the wave speed is $c = c_0/n$ where $n \sim 3.5$ is the far-infrared c-axis refractive index and $c_0$ the vacuum speed of light. Synchronization of the junctions and emission arise when the resonance condition $f_J = f$ is met. This mechanism of emission appears – at first sight – inconsistent with any tunability of the emitted radiation since the cavity frequency is determined by geometry and thus fixed. Even so, tunable THz-emission has been observed from mesas of various shapes [12-15] at very high bias.

Under these conditions highly inhomogenous mesa temperatures can arise in the self-heating regime inducing so-called 'hot spots', which can give rise to an effectively variable resonance cavity. Here, we present data in the low-bias regime were the cavity is set by the geometry of the sample.

We acquired detailed far-infrared spectra of the emitted radiation as function of bias voltage and temperature. Contrary to expectations, we demonstrate that the emission frequency can be tuned by as much as 12 % by varying bias voltage and temperature, respectively. We attribute the tunability to the fact that actual devices have lower symmetry than rectangular – as was implicitly assumed in previous work [2, 4, 7, 16]. Under these conditions the totally uniform synchronized state cannot exist, and for a non-uniform state the resonance condition is modified. Numerical simulations based on the Lawrence-Doniach model give a good account of our results.

A slightly under-doped Bi-2212 single crystal ($T_c$ = 76.3 K) was mounted on a sapphire substrate with conducting silver epoxy. Using optical lithography and argon ion milling, a cavity in form of a 300 x 80 x 1 micron mesa was patterned onto a freshly cleaved surface [2]. Measurements were performed with the device mounted in a continuous-flow cryostat. As per Figure 1, terahertz emission was collected from both sides of the mesa, respectively for measuring the overall intensity (using a chopper and lock-in amplifier at 77 Hz) and for measuring the emission spectrum using a Bruker Vertex 80V FT-IR spectrometer at a spectral resolution of 0.075 cm$^{-1}$. In both cases, the detector used was a liquid helium-cooled silicon bolometer. The mesa voltage was measured in a three-probe configuration (see Fig. 1a), with the bias current initially ramped up until all the junctions in the mesa were driven into the resistive state. The current was then swept down until emission intensity (above background) could be detected, and repeated spectra were acquired at a range of mesa voltages. The Bi-2212 mesa studied here has a sloping sidewall profile and trapezoidal cross-section, being wider at the bottom by as much as 20%. This shape is typically seen on Bi-2212 mesas [2, 5, 17]. As a result, biasing a current through the mesa will result in increasing DC current density - as well as junction voltage and Josephson frequency - from the bottom to the top of the stack. Depending on the shallowness of the sidewall profile, this spread in junction area may seriously compromise the phase coherence and power emission of the device, as in this regime the emission power is proportional to the square of the number of junctions oscillating in phase.

Fig. 2a shows the current-voltage (I-V) characteristics at 30 K and the simultaneously measured emission power (only data for decreasing current are shown). On increasing bias current the IV follows a series of quasi-particle branches [3] until it jumps to the McCumber branch on which the entire junction stack is in the resistive state. Emission occurs on decreasing current at mesa-voltages below 0.8 V. For mesas of this width and doping state, some of the junctions tend to re-switch into the zero-voltage state – as seen by the jumps in the IV-curve and emission power - before the voltage can be sufficiently reduced to fully trace the lower limit of the emission feature. Within the RCSJ model of a Josephson junction this retrapping voltage is given by the Josephson plasma frequency which is proportional to $(I_c \Phi_0 / C)^{1/2}$ [18, 19] and thus drops as $T$ is increased. Here, $C$ is the junction capacitance. Due to thermal fluctuations quantities such as the critical current, retrapping current and voltage are smeared as described by probability distributions [18-20]. Furthermore, in junction stacks the switching dynamics may be altered depending on the state of neighboring junctions [21]. Nevertheless, the retrapping voltage increases with decreasing temperature limiting the THz-emission at low temperatures. For the same reason, it is not possible to observe emission by approaching the resonance voltage from below for this particular mesa. While there is significant self-heating of the mesa at high bias current at which the I-V characteristics bends back on itself, this is negligible at the return-branch current levels at which emission is seen [22-24].

No emission is seen at bath temperatures of 50 K and above, while below 20 K, the junctions tended to retrap before the resonance voltage could be reached. As long as the entire mesa remains switched to the resistive state, the emission intensity, frequency, and linewidth characteristic with respect to voltage is highly reproducible at any given temperature. Fig. 2b shows the variation of the IV-curve and emission power in detail around the resonance for temperatures of 20 and 30 K. The power drawn by the cavity resonance can be evaluated from the I-V dependence as product of the excess current $\Delta I$ and the voltage V. It is difficult to determine the exact amount of this power, especially when the baseline I-V characteristic (due to quasiparticle tunneling in the absence of the resonance) can only be measured at voltages above the resonance, as is the case in Figure 2b. Nonetheless, the figure clearly shows this to be of the order of 20 μW at peak emission. By contrast, when allowing for the solid angle subtended by the collection optics, the total radiated power is closer to 0.5 μW, implying that most of the power delivered to the cavity resonance is re-absorbed in the chip - either in the mesa itself or in the bulk of the crystal - before it can be emitted. Furthermore, since the thickness $t$ of the mesa ($t$ ~1μm) is substantially smaller than the wavelength of the emitted radiation (~650 µm) the transmission coefficient through the side faces of the mesas is small. The data directly show the expected decrease of the re-trapping voltage with increasing temperature. Furthermore, the entire emission feature shifts to lower voltages at increasing temperature. This shift in voltage corresponds to shift to lower emission frequency at higher temperature as seen directly in the emission spectra shown in Fig. 3 for temperatures of 20 K and 30 K, respectively.

This temperature dependence of the emission feature arises from the asymmetric geometry of actual mesa devices. This asymmetry arises from a) the non-vertical side walls and b) different boundary conditions at the top (Au-contact) and bottom (contact to base crystal) of the mesa. Electromagnetic waves in strongly layered superconductors such as BSCCO exist as Josephson plasma waves. In a sample containing $N$ layers, $N$ different modes are expected for fixed in-plane wave vector $k$ [12, 25]. These are indexed by their wave vector perpendicular to the planes, $q$. The dispersion relation for these waves is given by $\omega^2 = \omega_{pl}^2 + c^2 k^2$ with $c^2 = c_0^2 / \varepsilon_c \left[ 2(1 - \cos(qs)) \lambda_{ab}^2 / s^2 + 1 \right]$. Here, $\omega_{pl} = c_0 / \sqrt{\varepsilon_c} \lambda_c$ is the Josephson plasma frequency, which for our samples is of the order of 70 GHz. $\varepsilon_c$ is the c-axis far-infrared dielectric constant of BSCCO, $\lambda_c$ and $\lambda_{ab}$ are the c-axis and $ab$-plane London penetration depths, and $s = 1.56$ nm is the repeat distance of the $CuO_2$-bilayers in BSCCO. In the resistive state the gap in the plasma wave spectrum, $\omega_{pl}$, is suppressed to zero. The mode with $q = 0$ is uniform along the c-axis and corresponds to in-phase oscillations of all the junctions in which case the wave speed reduces to the temperature independent value of $c = c_0/n$ as described above. The allowed values of $q$ are determined by the boundary conditions at the top and bottom surfaces of the mesa. Due to the high conductivity of the gold contact at the mesa top, it can be treated as an ideal conductor. This corresponds to an antinode for the oscillating c-axis electric field. The boundary condition at the interface between the mesa and the base crystal at the mesa bottom is more complicated [26]. Analyzing this condition, we concluded that with good accuracy the exact condition can be replaced by the simple approximation of vanishing c-axis electric field at this interface. Therefore, the lowest-order modes that are consistent with the actual mesa geometry would have $q = \pi m/2t$ with $m = 1, 3, ...$ . In analyzing the modes in BSCCO mesas several sets of boundary conditions have been invoked. For instance, for free standing mesas with highly conducting gold contacts on top and bottom [10] antinodes for the oscillating c-axis electric field at both surfaces arise leading to the set of allowed wave vectors $q = \pi m/t$, $m = 0,1,2...$ This geometry does allow for the uniform mode, m = 0, corresponding to a temperature independent mode frequency $f_0 = c_0/2nw$. In other simulations [12, 25] the uniform c-axis current condition was used which corresponds to the nodes of the oscillating electric field at both surfaces leading to the same set of allowed wave vectors but without the uniform mode. For our geometry, the dispersion relation of the Josephson plasma waves always yields a temperature-dependent wave speed of $c^2 = c_0^2 \big/ \varepsilon_c \left(1 + \left(m\pi\lambda_{ab}/2t\right)^2\right)$ and therefore a temperature-dependent frequency $f^2 = f_0^2 \big/ \left(1 + \left(m\pi\lambda_{ab}/2t\right)^2\right)$. As the penetration depth increases with increasing temperature the frequency decreases as is observed in the data. Since the emission frequency depends on temperature and bias voltage (see below) uncertainty arises in determining the temperature dependence quantitatively. However, choosing the center of the emission feature as indicated in Fig. 3 yields a change in frequency by ~8.5 % between 20 K and 40 K. The temperature dependence of $\lambda_{ab}$ has been determined from microwave cavity measurements for optimally doped BSCCO [27] and from ac-susceptibility measurements on powder samples with various doping levels [28]. Fig. 4 shows the measured temperature dependence of the average emission frequency together with the expected dependence determined by scaling the data for $\lambda_{ab}$ for the sample with doping level 0.124 from Ref. 26 by a factor of 1.5, giving $\lambda_{ab}(0) \sim$ 550 nm. This scaling may be reasonable since our crystal has $T_c \sim 76.3$ K whereas the sample from Ref. 26 has ~ 80 K. We normalize the frequency data to the value at 20 K, the lowest temperature for which retrapping did not preclude the experimental determination. The experimental temperature dependence is very well described by the variation of the $m = 1$ – mode. This observation is also consistent with the fact that the $m = 3, 5$ modes would require unexpectedly small values of the penetration depth in order to match the measured emission frequency.

We also notice that the emission peak amplitude varies quasi-periodically as a function of voltage, or indeed of frequency as is plotted in Fig. 3. At maxima of the amplitude, the observed emission linewidth is limited by – and possibly narrower than – the 0.075 cm$^{-1}$ spectrometer resolution (= 2.25 GHz). However, when the voltage is adjusted away from these maxima, the lines broaden until they have significant measureable width, and near minima of the peak amplitude, the lines may even become observably bimodal. This behavior is reversible with respect to bias current, is highly reproducible, and is in strong qualitative agreement with our numerical simulations (see below) which suggest that at certain bias voltages, the zone of radiating junctions becomes unstable and splits into two zones, each phase-locked to a different frequency.

We performed extensive numerical simulations of the collective behavior of stacks of intrinsic Josephson junctions in BSCCO with the goal to determine how the distribution of junction cross sections influences their synchronization and frequency tuning near the cavity resonance. Samples used in THz-emission experiments contain typically ~1000 junctions. Realistic simulations of such large systems are extremely time consuming, however, stacks of $N \sim 100$ junctions can be simulated in a reasonable amount of time. For our simulations we used the dynamic equations that describe the time evolution of the reduced c-axis electric field ($e_n$) in junction number $n$, the phase difference across junction $n$ ($\theta_n$), the in-plane phase gradient ($k_n$), and magnetic fields ($h_n$) [4, 10]

$$\frac{\partial e_n}{\partial \tau} = -v_c e_n - g(u)\sin\theta_n + \frac{\partial h_n}{\partial u} + \tilde{j}_z(u,n,\tau) \quad (1a)$$

$$\frac{\partial \theta_n}{\partial \tau} = e_n \quad (1b)$$

$$v_{ab}\frac{\partial k_n}{\partial \tau} = -[k_n + h_n - h_{n-1}] + \tilde{j}_{ab}(u,n,\tau) \quad (1c)$$

$$h_n = l^2\left(\frac{\partial \theta_n}{\partial u} - k_{n+1} + k_n\right) \quad (1d)$$

In these equations the unit of length is the Josephson length $\lambda_J$, the unit of phase gradients is $1/\lambda_J$, the unit of magnetic field is $\Phi_0/(2\pi\gamma\lambda_{ab})^2$, and the unit of the electric field $\Phi_0\omega_p/(2\pi c_o s)$. These reduced equations depend on three materials parameters, $v_c = 4\pi\sigma_c/(\varepsilon_c\omega_p)$, $v_{ab} = 4\pi\sigma_{ab}/(\varepsilon_c\omega_p\gamma^2)$ and $l = \lambda_{ab}/s$, where $\sigma_c$ and $\sigma_{ab}$ are the components of the quasiparticle conductivity and $\gamma$ is the anisotropy of the penetration depth. In our simulations we used $v_c = 0.01$, $v_{ab} = 0.2$ and $l = 50$ and 100, respectively. All quantities are assumed to be y-independent (long side of the mesas). To model thermal fluctuations and to improve equilibration, we include noise currents $\tilde{j}_z(u,n,t)$ and $\tilde{j}_{ab}(u,n,t)$, which are defined by the correlation functions

$$\langle \tilde{j}_z(0,0,0)\tilde{j}_z(u,n,\tau)\rangle = 2v_c\tilde{T}\delta(u)\delta(\tau)\delta_n \quad (2a)$$
$$\langle \tilde{j}_{ab}(0,0,0)\tilde{j}_{ab}(u,n,\tau)\rangle = 2v_{ab}\tilde{T}\delta(u)\delta(\tau)\delta_n \quad (2b)$$

The noise amplitude is determined by the effective temperature $\tilde{T}$. As our model is two-dimensional, this temperature cannot be related to the real temperature in a simple way. The equations are solved for stacks containing $N = 50 - 100$ junctions with a lateral size decreasing as function of the junction index as $L_n = L(1-\alpha n/N)$ corresponding to the trapezoidal-shaped mesas. We assume simple non-radiative boundary conditions at the edges, $k_n = 0$, $\partial \varphi_n /\partial u = \pi I /2 l^2$ at $u = 0, L_n$, where $I = j_n L_n$ is the total transport current.

Due to the distribution of the junction cross-sections, the stack is typically in the incoherent state in which junctions oscillate at different frequencies. In this state the local Josephson frequency $\omega_n$ is inversely proportional to the junction width $\omega_n = I/(L_n v_c)$. The transition to the coherent state may only occur when the average Josephson frequency $\omega_{av} \approx I/L_{mid} v_c$ approaches the cavity resonance frequency $\omega_r \approx l\pi/L_{mid}$, where $L_{mid} = L(1-\alpha/2)$ is the width of the junction in the center of the stack. This transition only happens when the inhomogeneity parameter $\alpha$ is sufficiently small. In the coherent state a strong cavity mode is excited which forces either all junctions in the stack or part of them oscillate at the same frequency. Three possible dynamic scenarios are illustrated in Fig. 5. When the parameter $\alpha$ characterizing the spread of junction widths exceeds some critical value there is no synchronization. The system does not notice the resonance voltage at all. This case is realized for $\alpha = 0.1$ in Fig. 5 row a). For smaller values of $\alpha$, the system experiences the synchronization transition, which may be seen as large excess current in the IV dependence [29]. We can see that at $\alpha = 0.0625$ (row c) the whole stack is synchronized while at $\alpha = 0.08$ (row b) the partially synchronized state is realized. In the latter state the voltage change with increasing current is mostly determined by junctions, which are not in the synchronized cluster leading to the formation of a "voltage gap" between synchronized and unsynchronized junctions. On the other hand, the oscillating electric fields of the cavity mode are excited in the whole stack, including the region of unsynchronized junctions. We observed strongly hysteretic behavior with respect to direction of current sweep. The synchronization is always much more pronounced at the increasing-current branch, as illustrated in Fig. 5c. The pronounced hysteresis has not yet been observed experimentally, however, the overall behavior obtained for *decreasing* bias resembles the experimental results. The dependences on decreasing bias current of the excess current and the mode amplitude together with visualizations of the junction voltages for two current values are presented in Fig. 6 for a stack with $N = 100$ and $\alpha = 0.8$. We found that a relatively small group of synchronized junctions is first formed near the bottom of the stack. In contrast to the strongly synchronized state, the number of junctions in the cluster and its locations varies with current leading to an irregular current dependence of the mode amplitude and excess Josephson current. This pattern bears a strong resemblance with our experimental data shown in Fig. 3 as well as with results previously obtained on a 60–µm mesa [17]. As one can see from the visualization of junction voltages, the synchronized bands moves up in the stack with decreasing current. At the same time, the resonance frequency decreases resulting in the voltage tunable emission frequency. As the bias voltage decreases, a zone of phase-locking develops and grows from the bottom of the mesa, where the junction area is largest, and hence $j_{DC}$ and $V_J$ are lowest, and $\omega_n$ is closest to the cavity resonance. With decreasing voltage, this zone progressively moves upwards, towards the narrower junctions at the top of the mesa.

In summary, we have demonstrated that in Bi-2212 mesa devices with sloped sidewalls, the emission frequency can be controlled via varying the bias voltage across the mesa. Significant further tunability can be achieved by varying the device temperature, making for a total frequency tunability in excess of 12%. The observed behavior of the device's terahertz emission with respect to varying both voltage and temperature are in qualitative agreement with predictions made by our numerical simulations. It would be useful in future to study devices with lower width in the same manner as done for this 80 x 300 micron device. A smaller width would shift the resonance to higher voltages, and allow us to study its lower limits without the junctions retrapping.

This research was funded by the Department of Energy, Office of Basic Energy Sciences, under Contract No. DE-AC02-06CH11357. We gratefully acknowledge Ralu Divan and Alexandra Imre of the ANL Center for Nanoscale Materials for assistance with lithography.

References

[1] B. D. Josephson, Phys. Lett. **1**, 251 (1962).

[2] L. Ozyuzer, A. E. Koshelev, C. Kurter, N. Gopalsami, Q. Li, M. Tachiki, K. Kadowaki, T. Yamamoto, H. Minami, H. Yamaguchi, T. Tachiki, K. E. Gray, W.-K. Kwok, U. Welp, Science **318**, 1291 (2007).

[3] R. Kleiner, F. Steinmeyer, G. Kunkel, P. Müller, Phys. Rev. Lett. **68**, 2394 (1992).

[4] A. E. Koshelev, L. N. Bulaevskii, Phys. Rev. B **77**, 014530 (2008).

[5] K. Kadowaki, H. Yamaguchi, K. Kawamata, T. Yamamoto, H. Minami, I. Kakeya, U. Welp, L. Ozyuzer, A. E. Koshelev, C. Kurter, K. E. Gray, and W.-K. Kwok, Physica C **468**, 634 (2008).

[6] H. Minami, I. Kakeya, H. Yamaguchi, T. Yamamoto, K. Kadowaki, Appl. Phys. Lett. **95**, 232511 (2009).

[7] M. Tsujimoto, K. Yamaki, K. Deguchi, T. Yamamoto, T. Kashiwagi, H. Minami, M. Tachiki, K. Kadowaki, R. A. Klemm, Phys. Rev. Lett. **105**, 037005 (2010).

[8] M. Tsujimoto, K. Yamaki, T. Yamamoto, H. Minami, K. Kadowaki, Physica C **470**, S779 (2010).

[9] S.-Z. Lin and X. Hu, Phys. Rev. Lett., **100**, 247006 (2008), Phys. Rev. B **79**, 104507 (2009); X. Hu and S.-Z. Lin, Supercond. Sci. Technol. **23**, 053001 (2010).

[10] A. E. Koshelev, Phys. Rev. B **78**, 174509 (2008).

[11] V. M. Krasnov, Phys. Rev. B **83**, 174517 (2011).

[12] H. B. Wang, S. Guenon, J. Yuan, A. Iishi, S. Arisawa, T. Hatano, T Yamashita, D. Koelle, R. Kleiner, Phys. Rev. Lett. **102**, 017006 (2009).

[13] H. B. Wang, S. Guenon, B. Gross, J. Yuan, Z. G. Jiang, Y. Y. Zhong, M. Grünzweig, A. Ishi, P. H. Wu, T. Hatano, D. Koelle, R. Kleiner, Phys. Rev. Lett. **105**, 057002 (2010).

[14] S. Guenon, M. Grünzweig, B. Gross, J. Yuan, Z. G. Jiang, Y. Y. Zhong, M. Y. Li, A. Ishi, P. H. Wu, T. Hatano, R. G. Mints, E. Goldobin, D. Koelle, H. B. Wang, R. Kleiner, Phys. Rev. B **82**, 214506 (2010).

[15] K. Kadowaki, M. Tsujimoto, K. Yamaki, T. Yamamoto, T. Kashiwagi, H. Minami, M. Tachiki, R. A. Klemm, J. Phys. Soc. Jpn 79, 023703 (2010); K. Yamaki, M. Tsujimoto, T. Yamamoto, A. Furukawa, T. Kashiwagi, H. Minami, K. Kadowaki, Optics Express **19**, 3193 (2011).

[16] R. A. Klemm, E. R. LaBerge, D. R. Morley, T. Kashiwagi, M. Tsujimoto, K. Kadowaki, J. Phys.: Condens. Matter **23**, 025701 (2011).

[17] K. E. Gray, L. Ozyuzer, A. E. Koshelev, C. Kurter, K. Kadowaki, T. Yamamoto, H. Minami, H. Yamaguchi, M. Tachiki, W.-K. Kwok, U. Welp, IEEE Appl. Superconductivity **19**, 886 (2009).

[18] M. G. Castellano, G. Torrioli, F. Chiarello, C. Cosmelli, P. Corelli, J. Appl. Phys. **86**, 6405 (1999).

[19] V. M. Krasnov, T. Golod, T. Bauch, P. Delsing, Phys. Rev. B **76**, 224517 (2007).

[20] E. Ben-Jacob, Phys. Rev. A **26**, 2805 (1982).

[21] P. A. Warburton, S. Saleem, J. C. Fenton, M. Korsah, C. R. M. Grovenor, Phys. Rev. Lett. **103**, 217002 (2009).

[22] C. Kurter, K. E. Gray, J. F. Zasadzinski, L. Ozyuzer, A. E. Koshelev, Q. Li, T. Yamamoto, K. Kadowaki, W.-K. Kwok, M. Tachiki, U. Welp, IEEE Appl. Superconductivity **19**, 428 (2009).

[23] C. Kurter, L. Ozyuzer, T. Prolier, J. F. Zasadzinski, D. G. Hinks, K. E. Gray, Phys. Rev. B **81**, 224518 (2010).

[24] A. Yurgens, arXiv: 1005.2932.

[25] R. Kleiner, Phys. Rev. B **50**, 6919 (1994); L. N. Bulaevskii, M. Zamora, D. Baeriswyl, H. Beck, J. R. Clem, Phys. Rev. B **50**, 12831 (1994); N. F. Pedersen, S. Sakai, Phys. Rev. B **58**, 2820 (1998).

[26] A. E. Koshelev and L. N. Bulaevskii, Journal of Physics: Conference Series **150** 052124 (2009).

[27] T. Jacobs, S. Sridar, Q. Li, G. D. Gu, N. Koshizuka, Phys. Rev. Lett. **75**, 4516 (1995); S.-F. Lee, D. C. Morgan, R. J. Ormeno, D. M. Broun, R. A. Doyle, J. R. Waldram, K. Kadowaki, Phys. Rev. Lett. **77**, 735 (1996).

[28] W. Anukool, S. Barakat, C. Panagopoulos, J. R. Cooper, Phys. Rev. B **80**, 024516 (2009).

[29] The excess Josephson current is defined as $\delta j(V) = j(V) - \Sigma_c V$ where $\Sigma_c$ is the total quasiparticle conductivity of the stack. The excess current is proportional to the amplitude of the excited cavity mode and serves as a convenient measure of the resonance strength.

Figure captions

Fig. 1: (Color online) (a) Diagram of the device structure on a Bi-2212 crystal (after [2]). Applying the correct bias current excites the transverse fundamental cavity mode on the width $w$. Terahertz radiation is emitted from the long side-faces with the highest intensity occurring roughly at 45 deg to the crystal surface [6]. (b) Arrangement of the measurement apparatus. Radiation emerges from the chip in two beams at roughly 90 deg relative to each other. One of these is used for detecting the overall emission intensity (Bolometer A) while the other is measured by the spectrometer, which employs Bolometer B as its detector.

Fig. 2: (Color online) (a) Current-voltage characteristics (red) and THz-radiation power (orange) for a 80 x 300 micron mesa. The hysteretic behavior of underdoped IJJs, the backbending due to self-heating at high bias currents (which can be weakly detected by the bolometer) and the progressive retrapping of the entire stack as the current is swept back down are seen. THz-emission is observed near mesa-voltages of ~0.7 V, 0.45 V and 0.33 V. The two low-voltage emission features result from subsets of junctions in the stack. (b) Detail of the main emission peak at 20 K (blue lines) and 30 K (red lines), showing excess current (and power) drawn by the mesa when the THz resonance is excited. The shift to lower voltage of the retrapping and of the emission feature with increasing temperature is seen.

Fig. 3: (Color online) FTIR emission spectra taken at a range of mesa bias voltages at (a) 20 K and (b) 30 K. With increasing temperature the emission band shifts to lower frequencies. The envelope of the emission peaks tracks well the voltage dependence of the total emission power as shown in Fig. 2b. The arrows mark the average frequencies used in Fig. 4.

Fig. 4: (Color online) Temperature dependence of the measured emission frequencies (red symbols) and of the calculated emission frequencies of the m=0 and m=1 modes using the penetration depth data from Ref. 16. The error bars are determined by the width of the emission range shown in Fig. 3. The inset shows schematics of the c-axis dependence of the cavity mode.

Fig. 5: (Color online) Illustration of three possible scenarios in stacks with a distribution of junction areas as characterized by the parameter $\alpha$. The left column of plots shows I-V characteristics and the right column shows the distribution of the average voltage drops for all junctions in the stack at the current values that are marked in the IV-curves by boxes. The parameters used in these simulations are $N = 50$, $l = 50$, $L_1 = 12.5$ and $\tilde{T} = 0$.

Fig. 6: (Color online) The current dependences of the resonance-mode amplitude (a) and the excess current (b) obtained by simulations of a stack with parameters $N = 100$, $\alpha = 0.8$, $l = 100$, and $\tilde{T} = 0.02$ on decreasing bias current. The non-monotonous voltage dependence of the mode amplitude agrees well with the measured voltage dependences shown in Figs. 2b and 3. Plots on the right show the distribution of the average junction

voltage $V_n$ for two bias currents. The outline of the colored areas represents the trapezoidal cross section of the mesa. The voltage values are also visualized by color coding revealing a band of synchronized junctions moving up in the stack with decreasing bias current.

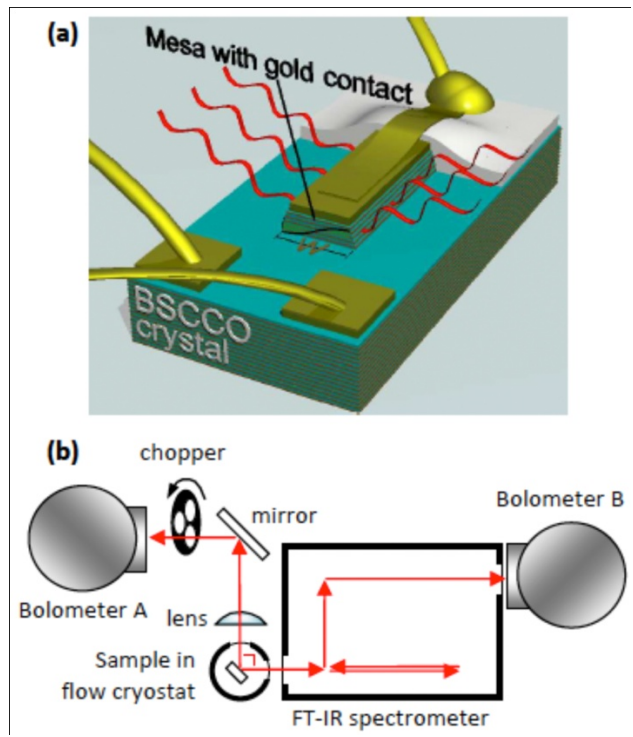

Fig. 1

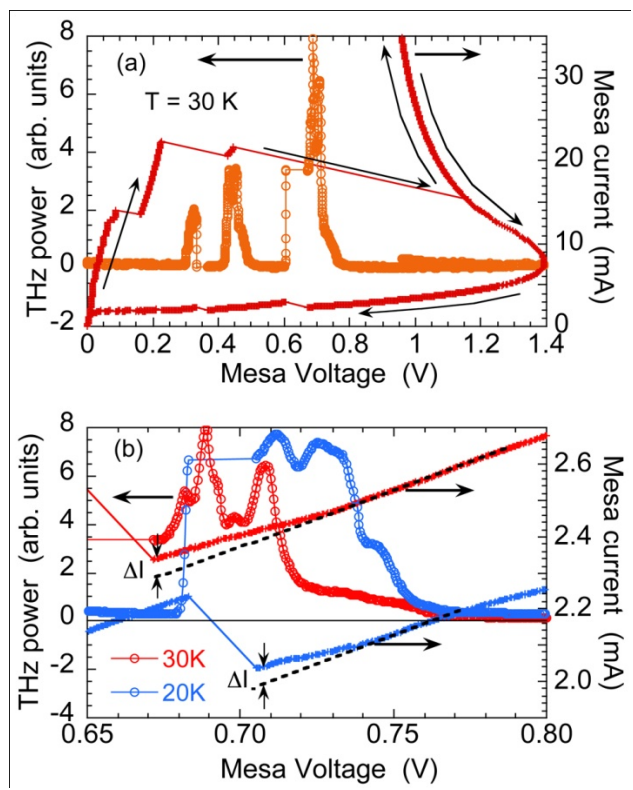

Fig. 2

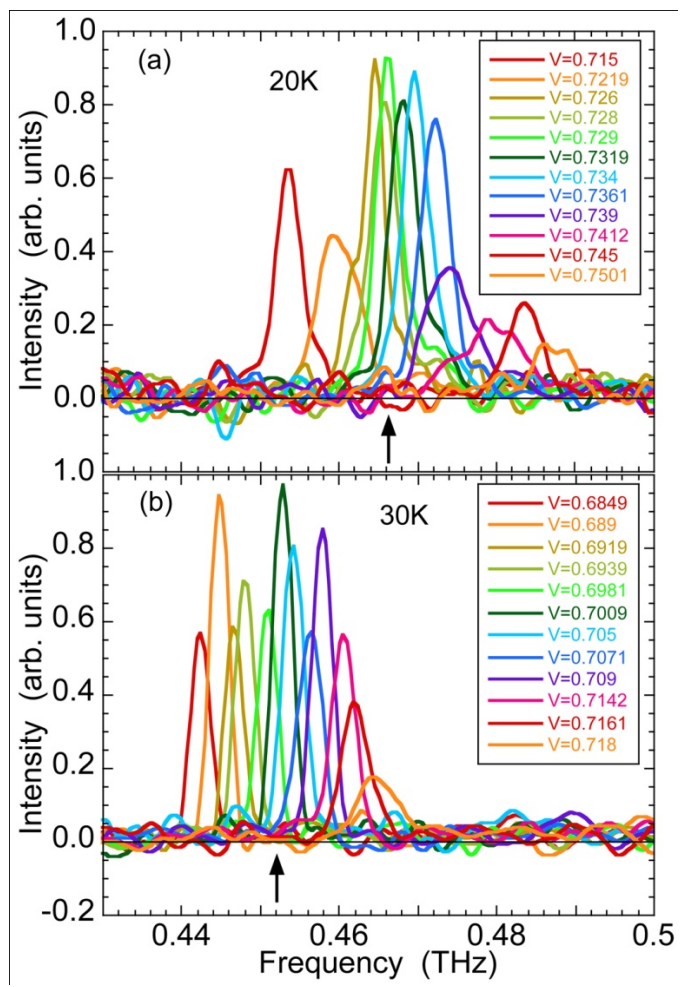

Fig. 3

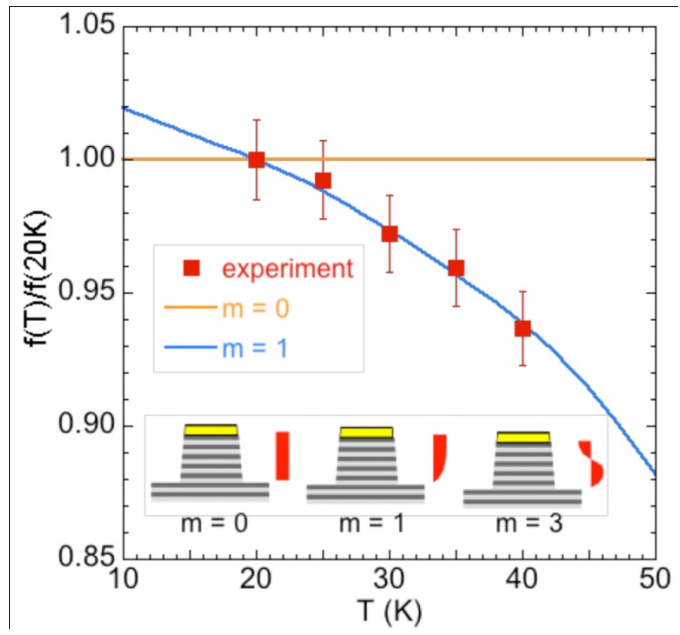

Fig. 4

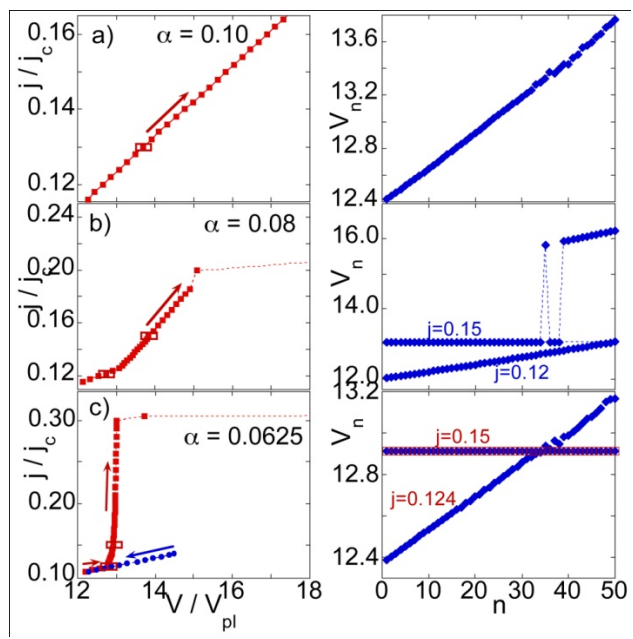

Fig. 5

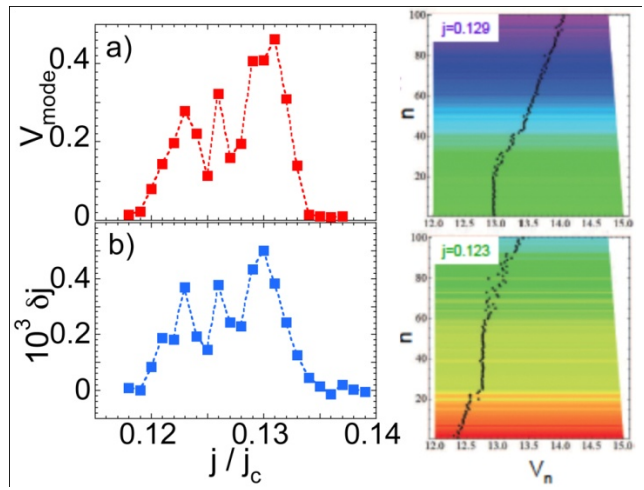

Fig. 6